\documentstyle[preprint,prb,aps]{revtex}
\newcommand{\be}{\begin{equation}}      
\newcommand{\ee}{\end{equation}}
\newcommand{\bea}{\begin{eqnarray}}     
\newcommand{\eea}{\end{eqnarray}}
\newcommand{\beb}{\begin{eqnarray*}}    
\newcommand{\eeb}{\end{eqnarray*}}
\newcommand{\D}{{\cal D}}
\renewcommand{\phi}{\varphi}
\newcommand{\Cinv}{C^{-1}}

\renewcommand\phi{\varphi}
\newcommand\Real{{\,\,\rm Re\,}}
\newcommand\half{{\displaystyle 1\over\displaystyle 2}}
\newcommand\uvec{{\bf u}}
\newcommand\vvec{{\bf v}}

\newcommand\pvec{{\bf p}}
\newcommand\rvec{{\bf r}}
\newcommand\xvec{{\bf x}}
\newcommand\Xvec{{\bf X}}

\newcommand\Avec{{\bf A}}
\newcommand\bvec{{\bf b}}
\newcommand\evec{\hat{\bf e}}

\newcommand\grad{{\bf\nabla}}
\newcommand\dst{\displaystyle}

\begin{document}
\title{Geometrical Defects in Josephson Junction Arrays}

\author{Kieran Mullen}
\address{ University of Oklahoma, Department of Physics,
             Norman, OK 73019-0225 
}

\maketitle

\begin{abstract}
Dislocations and disclinations in a lattice of Josephson junctions will 
affect the dynamics of vortex excitations within the array.
These defects effectively distort the space in which the excitations
move and interact.   
We calculate the interaction energy between such defects and
excitations, determine vortex trajectories in twisted lattices,
and discuss the consequences for experiment.
\end{abstract}


\input epsf.sty

\section{Introduction}
\label{sec:intro}

A host of interesting experiments have already been performed on regular
arrays of Josephson junctions.  
Tuning the ratio of the pair charging energy to the Josephson
coupling energy in such experiments
allows one to pass from a regime where the dynamics is
dominated by vortex excitations, to one where it is dominated by 
charge soliton excitations.\cite{schon}
In the vortex dominated regime one can examine such issues as
the ballistic propagation
of vortices,\cite{vor-dyn1,vor-dyn2}
the Kosterlitz-Thouless (KT) vortex unbinding
transition,\cite{KT-vor1,KT-vor2,KT-vor3} 
and how the interference of vortices is influenced
by electric charges in the Aharanov-Casher effect.\cite{casher-exp}
In the charge dominated regime the parallel issues of the propagation
of solitons,\cite{soliton-us,soliton-them} the possibility of
the KT charge unbinding transition\cite{soliton-them}
 and how the
interference of charges is influenced by magnetic fields in
the Aharonov-Bohm effect\cite{ab-1,ab-2} have been 
investigated.  The charges and vortices in a superconductor
are approximately dual, and the physics of the two  reflect this
symmetry.\cite{fisher,fazio}

In this paper we study the motion of vortices in arrays that possess
topological defects.
Studies of Josephson junction arrays usually focus on  either square or 
trangular lattices, where the structure is uniform.  It is an
unspoken assumption that the lattice of junctions be free from
{\it spatial} defects such as dislocations ({\em e.g.}
a missing line of junctions as in fig.(\ref{defects}.a)
or disclinations ({\em e.g.}
a missing wedge of junctions, as in fig.(\ref{defects}.b)).
We focus on the regime where $E_J$, is larger or on the order of 
$E_C$, so that vortices can be viewed as stable massive particles with
a mass determined by their charging energy.
Under these conditions vortices are free to move ballistically in the
array.
 We find that the
defects act as distortions  in the 2D world in which the excitations
move,  with two effects.  First, 
 ``straight lines'' (geodesics)  in this twisted coordinate system 
differ from those in Cartesian coordinates, 
and second the fields surrounding the
excitations are perturbed by the twisted geometry, alterring the
interaction between vortices and leading to an
interaction between vortices and the geometrical defects.

The fundamental origin of the interaction between vortices and topological
defects is not hard to see.  The 
self-energy of the vortex depends upon spatial gradients in the
superconducting phase.
Dislocations and disclinations involve either
the removal or insertion of extra junctions in the pristine lattice.
From the point of view of the vortices this means there are either
missing or added regions of space in  the two dimensional plane.  
This affects the gradients and thus the self energy.

In section \ref{sec:contdesc} below  we  
develop a continuum description of the dynamics on
a distorted lattice.  We then
analyze the effect of topological irregularities on the dynamics of a
single vortex in an array in which $E_J\geq E_C$, i.e., an array in
which a vortex can be viewed as a stable massive particle. 
In section \ref{sec:potential} we calculate the interaction of vortices or
charges on the twisted lattice. 
We use these results in section \ref{sec:motion} to calculate the motion of
vortices in a lattice with defects.  
The effects of the defects
should be observable in experiments in which the ballistic motion of
vortices is probed. 
We conclude in section
\ref{sec:conclusion}.

In the calculations below we use Einstein summation notation in which we
sum over repeated indices.  Roman indices range over the values 1 and 2,
greek indices for the  2+1 dimensional space-time range over 0, 1 and 2.
Bold face variables are 2-vectors; hatted vectors are normalized.

\section{Development of the continuum description}
\label{sec:contdesc}

We consider an array of Josephson junctions in an externally applied
magnetic field perpendicular to the array.  We assume that the array has
only smooth distortions in its structure, and a small number of defects
such as the dislocation or disclination in fig.\ref{defects}.  
Arrays of Josephson junctions are customarily described in terms of the
set of variables $\phi_a$, and $n_a$, the phase of the super--conducting order
parameter and the number of excess Cooper pairs 
on  super--conducting island $a$.
Although the islands themselves are extended objects, we consider the phase
and charge
as being defined at a discrete set of points.  (That is, we require 
the island width to be smaller than  or on the order of
the correlation length, so that the phase across the island is a constant). 
The thermodynamics of the
array is derivable from a Hamiltonian  made out of two contributions, the
Coulomb charging energy and the Josephson coupling energy. The partition
function can be written as a path integral over the charge and phase
degrees of freedom:
\be
{\large \cal Z} = \sum_{n_a}   \int \D\phi_a  \exp\left\{
 - {1 \over \hbar} S[n_a, \phi_{a}]\right\}
\ee
with the action given by,
\be
 S[n_a, \phi_a]= 
\int_0^\beta d\tau \left[ -i \hbar n_a \dot\phi_a 
- 2e^2 n_a \, n_b\,  \Cinv_{ab} 
 + \sum_{a,b }' E_j \left(1 - \cos(\phi_a-\phi_b -  \theta_{ab})
\right)\right]  \label{action}
\ee
where repeated indices are summed and the path integral in $\tau$ is
computed in the standard step-wise fashion.  The quantity
$\Cinv_{a,b}$ is an element of the inverse
capacitance matrix element
between islands $a$ and $b$, 
 $E_J$ is the Josephson coupling energy between two nearest neighbor
junctions,  and
$\beta= 1/k_B T$ where $T$ is the temperature and $k_B$ is Boltzman's
constant.  
The primed sum indicates that only nearest-neighbor phase differences are
included.
 The phase difference $\theta_{ab}$ is caused by the externally applied
magnetic field and is determined from the
$\Avec$, the magnetic vector potential,
via
\be
\theta_{ab}\equiv \int_a^b \Avec(\xvec)\cdot d\xvec .
\ee
where we have absorbed a factor of $2e/c$ into $\Avec(\xvec)$

We next make the standard Villain approximation to decompatify the 
phase.\cite{{villain}}
We introduce a set integer variables, $v^{(0)}_a$ defined on each island,
and $v_{ab}$, associated with each nearest-neighbor pair of islands
and approximate  eq.(\ref{action}) as:
\bea
Z \approx  c_0  \int \D n_a \int \D \phi_a \sum_{v_{ab}, v^{(0)}_a}
\exp\left\{ - {1\over \hbar} \int_0^\beta \left[
-in_a(\hbar\dot\phi_a + 2\pi v^{(0)}_a) - 
2e^2 \,\, n_a\,\,  n_b\,\,  \Cinv_{ab}\right.\right. \nonumber \\
 + \left.\left. \sum_{a,b}' {E_j \over 2 \hbar} 
\left( \phi_a - \phi_b - \theta_{ab}
+ 2\pi v_{ab}\right)^2\right]\right\} \label{act2}
\eea
where $c_0$ is a constant prefactor whose value depends upon $E_J$ and the
capacitance.  The variable $n_a$ is now continuous; by summing over
$v^{(0)}_a$ we recover only its integer contributions.  
Following the Villain transformation, the phase $\phi_{ab}$ are all regular
({\it i.e.}\/ non-compact) variables.  The vorticity is all contained in
the variables $v_{ab}$.

The system  described by 
eq(\ref{action}), or eq(\ref{act2}), is evaluated on a lattice of points, a
lattice that may possess topological
defects such as dislocations or disclinations.  
As in the case of a regular lattice, the long wavelength low
frequency properties are obtained from a continuum description of the
lattice.\cite{stern,fazio}
In the case of a distorted lattice, however, the continuum
description should reflect the lattice distortions.  
Our goal is the construction of a continuum Lagrangian
whose long wave length low frequency properties are the same as those
of the discrete model of eq(\ref{act2}).  

Although the phase $\phi_a$ is
only defined at a discrete set of points, we may define a continous
function $\phi(\xvec)$ which smoothly interpolates between the phases
defined at each lattice point.  We replace the phase difference between two
islands by a Taylor expansion:
\bea
\phi_a-\phi_b &=& \phi(\xvec_a) - \phi(\xvec_b) \nonumber \\
&\approx& {\bf\grad} \phi(\xvec_a) \cdot \uvec_{(i)}(\xvec_a)
\eea
where $\uvec_{(i)}(\xvec_a)$ is a vector connecting two adjacent islands.
There will be two such vectors ($i=1,2$) associated with each island, as
shown in fig.\ref{lattice}.  
The direction and length of the vectors will change as we
move through the distorted lattice.  In a similar fashion we associate a
pair of the  integers, $v_{ab}$ and $v_{ab'}$ with  these directions so
that 
$\vvec_{(1)}(\xvec_a)=  v_{ab}\,\,  \hat\uvec_{(1)}(\xvec_a) /
        u_{(1)}(\xvec_a)$  and
$\vvec_{(2)}(\xvec_a)=  v_{ab'}\,\,  \hat\uvec_{(2)}(\xvec_a) /
        u_{(2)}(\xvec_a)$
where $u_{(i)}(\xvec_a) \equiv | \uvec_{(i)}(\xvec_a)|$, and 
$\hat \uvec_{(i)}(\xvec_a) \equiv \uvec_{(i)}(\xvec_a)/ u_{(i)}(\xvec_a)$

Similarly, we approximate the contribution from the magnetic field by:
\be
\theta_{ab} \approx \Avec (\xvec_a) \cdot \uvec_{(1)}(\xvec_a)
\qquad \qquad 
\theta_{ab'} \approx \Avec (\xvec_a) \cdot \uvec_{(2)}(\xvec_a)
\ee
so that our partition function becomes
\bea
Z &\approx& c_0 \int \D n(\xvec_a) \int\D\phi(\xvec_a)
\sum_{\vvec_{(i)}(\xvec_a), v^{(0)}(\xvec_a)}
\exp\left\{ - {1\over \hbar} \int_0^\beta 
\left[ -i\, n(\xvec_a) (\dot\phi(\xvec_a) + 2\pi v^{(0)}_a) - 
\right.\right. \nonumber \\ 
&\phantom{=}& 
\left.\left. 2e^2 \,\,n(\xvec_a)\,\, n(\xvec_b)\,\, \Cinv_{\xvec_a, \xvec_b}
 +  \sum_{a} \sum_{i=1,2} {E_j \over 2 \hbar} 
\left[\left( \partial_k \phi(\xvec_a) -  A_k(\xvec_a) + 
2\pi v_{(i)}^k(\xvec) \right)u_{(i)}^k(\xvec)\right]^2\right]\right\} 
\label{act4}
\eea
We would like to turn this discrete sum
over the lattice points $\xvec_a$ and nearest neighbor vectors 
$\uvec_{(i)}(\xvec)$ into an integral over the Cartesian coordinates
$x_1$ and $x_2$ in the plane.  

All geometrical information about the lattice is contained within the
vectors $\uvec_{(i)}(\xvec)$, since they indicate the direction and
distance to the nearest neighbors.  
Locally each of the $\uvec_{(i)}(\xvec)$  can be written
as
\be
\uvec_{(i)} = \left(a^i_{\phantom{i}1} \,\, \evec_1 +a^i_{\phantom{i}2} \,\, \evec_2
\right) 
\ee
where we have suppressed the spatial 
dependence and the $\evec_i$ are the
Cartesian unit vectors.   
We note that 
the $a^i_{\phantom{i}j}$ are
the 2D analogs of basis triads in differential geometry;  they relate
differentials in the local, twisted reference frame (the $\uvec_{(i)}$) to
the Cartesian frame.\cite{K2}
We may rewrite gradient-squared terms as:
\be
\sum_{i=1,2}  \left(\grad \phi\cdot \uvec_{(i)}\right)^2 =
 \partial_j \phi \,\partial_k \phi \,\, 
 a_{i}^{\phantom{i}j} a_{i}^{\phantom{i}k}  
\ee
We introduce the matrix $g^{jk}\equiv a_{i}^{\phantom{i}j}
a_{i}^{\phantom{i}k}$; its inverse, denoted by $g_{jk}$, is the
{\it metric tensor}.  A small displacement in
the Cartesian  coordinates,  $d\xvec = dx^1 \,\, \evec_1 + dx^2 \,\, \evec_2$, 
has a distance when measured in the number of junction interfaces
crossed (or dimensionless junction ``hops'') of
\be
ds^2 = g_{jk} \,\, dx^j \,\, dx^k
\ee
whence the name ``metric''.

Furthermore, 
the area $a$ of the
plaquet bounded by $\uvec_{(1)}$ and $\uvec_{(2)}$ is
just 
\bea
{\cal A}&=& |\uvec_{(1)} \times \uvec_{(2)}| \\
  &=&  \sqrt{\det{g_{ij}}}  \,\, u_{(1)}\,\, u_{(2)} \\
  &\equiv&  g^{1/2} \,\,   u_{(1)}\,\, u_{(2)}
\eea
Note that the metric tensor and its determinant may vary in space.

We must also convert the capacitance terms to a continuum form.
 The original capacitance matrix satisfies
$Q_a = C_{ab} V_b$ where $V_b$ is voltage on island $b$;
if we assume only nearest neighbor coupling, then
this can be written as:\cite{fazio}
\be
(4C_1+C_0) V_a 
- C_1 \sum_{\rm neighbors}  V_b  = 2e \, n_a
\ee
where the sites $b$ are neighbors of $a$.
The variables
$C_0$ and $C_1$ are the diagonal and off-diagonal elements of the 
capacitance matrix.  
If we neglect
the diagonal self-capacitance,
($C_0=0$), it states that the discrete divergence
of $V$ equals the charge on a given island.
Through an analysis similar to what was done for the phase, we can define a
continuous field $V(\xvec)$ and replace the above discrete equation by
\be
(C_1 \,\, g^{-1/2} \partial_i g^{1/2} g^{ij} \partial_j + 
\tilde C_0(\xvec) )\,\,V(\xvec) = 2e\,\, n(\xvec)
\ee
where $n(\xvec)$ consists of a sum of $\delta$-functions, and $\tilde
C_0(\xvec)$ is a capacitance density given by $C_0/a$.
If we neglect this self-capacitance term, then the left hand side is
the Laplace-Beltrami (LB) operator, and 
the inverse capacitance matrix is equivalent 
 the Green function of this differential operator in the appropriate
continuum limit.
  Let us denote this Green function by
 $\Cinv(\xvec,\xvec')$ and define $U(\xvec,\xvec')=\lim_{C_0\to 0}
\Cinv(\xvec,\xvec')$.  In a flat space 
$\Cinv(\xvec,\xvec')=K_0(2 e \sqrt{C_1/\tilde C_0} \,|\xvec-\xvec'| )$ where 
$K_0(x)$ 
is the 
zero order Bessel function of imaginary argument, and 
$U(\xvec,\xvec')= (2\pi C_1)^{-1} \log{|\xvec-\xvec'|}$.\cite{K2}

If we take the continuum limit where $a$ becomes an infinitesimal  then we
can replace the sums by integrals over area and write the action as:
\bea
S[n(\xvec), \phi(\xvec)] &=& 
 - {1\over \hbar} \int_0^\beta \int d\xvec g^{1/2}(\xvec) 
\left[ -i\, n(\xvec) \left(\dot\phi(\xvec) + 
2\pi v^{(0)}(\xvec)\right) - \right. \nonumber \\
&\phantom{=}& +   {E_j \over 2 \hbar} 
\left( \partial_j \phi(\xvec) -  A_j(\xvec) + 2\pi v_{(i)}^j(\xvec) \right) 
\left( \partial_k \phi(\xvec) -  A_k(\xvec) + 2\pi v_{(i)}^k(\xvec) \right)
 g^{jk} 
\nonumber \\
&\phantom{=}& \left. 2e^2 \int d\xvec'\,\, g^{1/2}(\xvec')
\,\,n(\xvec)\,\, n(\xvec')\,\,\Cinv(\xvec,\xvec') \right] 
\label{act5}
\eea
where $n(\xvec_a)$ is the charge density.

We now wish to perform a series of operations on the action so as to
bring the vortex degrees of freedom to the fore.  These steps are standard
ones for deriving the interactions of vortices in 2D systems, made slightly
more tedious by the necessity to keep track of the local metric.  
While they appear messy and involved, one should keep in mind that these
steps are identical to what one would do if we 
had a regular, uniform array, but
wished to evaluate all quantities in curvilinear ({\it i.e.} polar)
coordinates.  Furthermore, 
if we choose our lattice to be smooth and undistorted,
so that $g_{ij}= \delta_{ij}$, we should  regain the standard vortex
interaction.

We first decouple the Josephson term via a Hubbard-Stratonavich
transformation, introducing a new field $\pvec(\xvec)$.  The second term then
becomes:
\be
\int d\xvec \,\,g^{1/2} \,\, \left[{1\over 2 E_j} \,\, p_i p_j g^{jk} +
(\partial_k\phi-A_k+2\pi v_{(i)}^k ) p_j g^{jk} \right]
\ee
The action is now linear in $\phi$; integrating it out we obtain the
constraint:
\be
\dot n(\xvec) = - g^{-1/2} \,\,\partial_k g^{1/2} g^{jk}\,\, p_j  \label{cont}
\ee
The right hand side is just the divergence of $\pvec$ expressed in
invariant form.  In order to satisfy this constraint we first
express it in a  ``space-time'' 
notation, using greek indices $\mu\in \left\{0, 1, 2\right\}$,
  setting $\partial_0=\partial_t$ and defining the 3-momentum
$p_\mu=(n(\xvec), p_1(\xvec), p_2(\xvec))$.
Then eq.(\ref{cont}) can be rewritten
\be
g^{\mu\nu} \,\, D_\mu p_\nu = 0  \label{div}
\ee
where $g_{\mu\nu}$ is the the 3x3 metric,
\be
g_{\mu\nu}\equiv 
\left( \begin{array}{c c}
1 & 0 \\
0 & g_{ij} 
       \end{array}\right)
\ee
and $D_\mu$ is the covariant derivative.  
Note that trivially $\det{g_{\mu\nu}} = \det{g_{ij}}$.   Eq.(\ref{div}) can
be satisfied if $p_\mu$ is the covariant curl of a vector field:
\be
 \epsilon^{\kappa\lambda\mu} g_{\nu \kappa} D_\lambda K_\mu
= 2 \pi p_\nu
\label{soln}
\ee
where $K_\mu$ is an auxiliary or gauge field and 
$\epsilon^{\kappa\lambda\mu}$ is the totally antisymmetric
Levi-Civita tensor.  Since covariant derivatives
of the metric are zero, eq.(\ref{div}) is satisfied automatically.
We rewrite our action in terms of the gauge fields $K_\mu$,
expressing  $n(\xvec)$ and $\pvec(\xvec)$  using eq.(\ref{soln})
As is common with such fields, we may choose a gauge condition for $K_\mu$
that will simplify the algebra: we choose
\be
g^{ij}D_i \,K_j = 0
\ee
in which case we the action becomes:
\bea
S[K_\mu] &=& -{1\over \hbar}\int d\tau \int d\xvec \,\,g^{1/2}(\xvec)\left[
 i \epsilon^{\mu\nu\lambda} 
( v_\lambda D_\mu K_\nu-A_\lambda \, D_\mu  K_\nu)
\right. \nonumber \\ 
&\vphantom{=}&  
+{1\over 8 \pi^2 E_J}
\left(\partial_i  K_0\,\, \partial_j K_0 +
\partial_0 K_i \,\, \partial_0 K_j \right)\,\, g^{ij} \nonumber \\
&\vphantom{=}&  \left.
+{e^2\over 2 \pi^2 C_1} \int d\xvec' g^{1/2}(\xvec') 
\epsilon_{ij}\epsilon_{k\ell} \,\,
D_i K_j(\xvec) \,\,  D_k K_\ell(\xvec') \,\, \Cinv(\xvec,\xvec')\right]
\eea
where 
 $A_\mu = (0, A_1, A_2)$.  Integrating by parts on the first term we
obtain
\bea
S[K_\mu] &=& -{1\over \hbar}\int d\tau \int d\xvec g^{1/2}(\xvec)\left[
 i  g^{\mu\nu}( J_\nu - \Phi_\nu) \,\, K_\mu
+{1\over 2 E_J}\,\, g^{ij}
\left( \partial_i K_0\,\,\partial_j K_0 +
\partial_0 K_i\,\,\partial_0 K_j \right)
\nonumber \right. \\ &\phantom{=}&\left. 
+\int d\xvec' g^{1/2}(\xvec') 
\epsilon_{ij}\epsilon_{k\ell}
D_i K_j(\xvec)\,\, D_k K_\ell(\xvec')\,\,\Cinv(\xvec,\xvec')\right]
\label{contact}
\eea
where $\Phi$ is a magnetic field density,  $\Phi_\mu=(B_z/\Phi_0, 0,0)$, 
and $J_\nu$ is a vortex current density related to our field
$\vvec$ by a covariant curl:\cite{{stern}} 
\be
J^\lambda = \epsilon^{\lambda\mu\nu } D_\mu \, v_\nu
\ee
This is the continuum version of the distorted Josephson junction lattice.
The $K_\mu$ field mediates the interation between vortex charges and
currents.   The action is quadratic in these fields, and thus the path
integrals over $K_\mu$ can be performed.   The resulting action will 
give the vortex Hamiltonian in the curved space.  This is done in
section \ref{sec:potential} below.

\section{Vortex Interactions in a Curved Space}
\label{sec:potential}

 The continuum action of eq.(\ref{contact}) is expressed in terms of the
gauge fields and the vortices.  We would like to integrate out the
$K_\mu$ so as to have an action solely in terms of the vortex degrees of
freedom.  First, we
 write the vortex current density as a sum over discrete point 
vortices located at positions $\Xvec^{(n)}$:
\be 
J_\nu= \sum_n \rho(\Xvec^{(n)}(t) ) \,\,\,\, \delta(\xvec-\Xvec^{(n)}(t)) \,\,
\,\,\left( 1, \dot X_1^{(n)}(t), \dot X_2^{(n)}(t)\right)
\ee
where $ \rho(\Xvec^{(n)})$ is the charge of the
vortex located at $\Xvec^{(n)}$.

We can integrate out the $K_0$ field as follows:  integration by parts on the
quadratic term yields $K_0 g^{-1/2} \partial_i g^{1/2} g^{ij} \partial_j
K_0$;   this combination of metrics and derivatives
 is just the LB operator, 
the generalization of the Laplacian
to this curved space.  Integrating out $K_0$ will yield
 a potential interaction
between vortices, $U(\xvec,\xvec')$.

We next examine the $K_i$ dependence.
The $\partial_t K_i$ terms correspond to a retarded interaction
between the vortices;  we can neglect this non-locality in time so long as
the characteristic velocity of vortices is much less than $\omega_j \ell$,
where $\omega_j=\sqrt{8E_JE_C}/ \hbar$, and $\ell$ is a characteristic
island size.\cite{fazio}  Doing so amounts to making the charge and
vortex degrees of freedom exactly dual.  We further assume that 
$C_0$ is small and we can approximate the inverse capacitance kernel
with $U(\xvec,\xvec')$.
As discussed above, $U(\xvec,\xvec')$ is the inverse of
the LB operator; 
when we perform the path integral it must cancel out the
powers of $\grad$ acting on $K_i$.  
This integration will generate two types of
terms quadratic in the vortex velocities.  The first will be self-interaction
terms that generate a mass for the vortex; the second terms are a velocity
dependent interaction between the vortices.\cite{{fazio}}
Our final action then depends 
only on the vortex degrees of freedom

Putting this all together we get:
\bea
S_{\rm eff}&=& {1\over\hbar} \int d\tau  \half  m_{\rm eff}\sum_n 
g_{ij}\,\,\dot X_i^{(n)} \,\,  \dot X_j^{(n)}  +  
\half \sum_{m,n} g_{ij}\,\,\dot X_i^{(m)} \,\,  \dot X_j^{(n)}
\,\,U(\Xvec^{(m)}, \Xvec^{(n)})
\nonumber \\
  & & 
+\half \sum_{m,n} \rho(\Xvec^{(m)})\,\,\rho(\Xvec^{(n)})
 \,\,U(\Xvec^{(m)}, \Xvec^{(n)})
\eea
where $m_{\rm eff}= \pi^2 \hbar^2 C_1/2e^2$  and we have suppressed the
explicit dependence on time.
The first term is the kinetic energy of the vortex in the twisted space;
the second term is the veloctity dependent
vortex-vortex interaction, and the third is vortex-vortex potential.


In order to demonstrate the effect of the defects, we calculate the 
interaction kernel $U(x,x')$ for 
 two simple cases, that of the disclination
and the dislocation.  In order to do so, we 
are simply solving for how point charges interact
in the 2D curved space defined by the distorted lattice. 
In flat space the field about a point charge falls off logarithmically.
If we set $z_1=x+i\,y$,
the 2D potential of a point charge located at $Z$ in a {\it flat} plane
can be written as:
\be
V(x,y)= V(z)= {1\over 2\pi} \Real \, q \ln(z-Z).
\ee
We will need this formulation below.  The interaction energy for a set of
charges in flat space is then:
\be
E_{\rm flat} = {1\over 4\pi} \sum_{i,j}  
\rho(\Xvec^{(i)}) \,\, \rho(\Xvec^{(j)}) \,\,
\log{\vert \Xvec^{(i)} - \Xvec^{(j)}\vert}
\ee
In this result we have neglected a constant
self-energy term for the vortices that
diverges as the log of the system size. This term either forces the net
vorticity of the system to be zero, or is compensated by the externally
applied field.

Although this is a continuum model for a problem on a lattice, it is known 
that 
this Green function is well approximated by that of the continuum 
problem.\cite{spitzer} This gives us confidence in similar
appoximations we make when we examine
 lattices with defects.

\subsection{Disclinations:}
\label{subsec:disc}

A negative disclination (fig(\ref{defects}b))  in 2D can 
be viewed as the projection of a cone into the
plane.  In order to form the cone, 
a wedge of sites is removed from the lattice. Next, the 
exposed edges are brought together, puckering the surface into a cone.  
Finally, the lattice is projected from  the 3D cone back down to a 2D plane.
This final distortion or stretching of the lattice is irrelevant from the
point of view of the excitations, since it does not change the 
number or connectivity of the junctions.  We can thus approximate 
the interaction energy
of a charge or vortex  excitation with a negative disclination
by calculating their energy on a cone, or equivalently, on a plane with
a wedge removed (fig.\ \ref{wedge}).
We therefore need only solve the Poisson equation for a single point
charge on a cone in order to obtain the potential $U(\xvec,\xvec')$.

This  problem on a cone is equivalent to 
a 2D electrostatics problem of a single
point charge on the planar surface with a missing wedge,
as depicted in fig.\ \ref{wedge}.  The boundary
condition across the cut is that the electric field is continuous.
We are free to choose our cut in any direction;  we choose to place
our charge on the positive x-axis and center the cut out wedge
along the negative x-axis.  The boundary condition now simplifies
to the requirement that the normal component of the field is zero on the
edges of the wedge, and the parallel component is continuous.  We can 
achieve this by a simple conformal mapping.\cite{panofsky}
We can open a wedge in the plane via the mapping $z\to z' = z^{1/p}$
For a wedge with an opening angle $\alpha$ (fig.\ \ref{wedge}) we require
$p=2\pi/(2\pi - \alpha)$.  Then 
\be
V(z) = \Real \ln(z^p - Z^p)  \label{cmap}
\ee
From this result we can obtain the vortex-vortex interaction energy.
In addition, the self-energy term generates 
an interaction energy between the charge and the disclination.
This  can
be found by calculating the electric field near the charge, subtracting 
the part due to the self-interaction, and integrating the force on
the charge to get
an effective potential.  Putting the two together we find that the
interaction energy of a set of vortices (or charges) on a lattice with a
disclination at the origin is: 
\bea
\label{vdisc}
E_{\rm disc}&=& {1 \over 4\pi } \left\{
 \sum_{i,j} \rho(\Xvec^{(i)}) \,\,\rho(\Xvec^{(j)})\,\, \right. \nonumber \\
 & & \log\left[|\Xvec^{(i)}|^{2p} + |\Xvec^{(j)}|^{2p} -
2|\Xvec^{(i)}|^{p}\,\,|\Xvec^{(j)}|^{p}\,\,\cos{p(\gamma_i-\gamma_j)}\right] +
\nonumber \\
& & \left. \sum_i \rho(\Xvec^{(i)})^2 \,\,(p-1)  \,\,\ln{X^{(i)}} \right\}
\eea
with $\gamma \equiv \arctan(X_2/X_1)$.
Eq. \ref{vdisc}
reduces to the correct result for $p=1$ (no distortion) and
$p=2$ (a half plane with an image charge at $-z_0$).  Translation
invariance is explicitly broken by the defect.

We can see a intuitively why such the electrostatic charge interacts
with the defects by considering the density of field lines.  Imagine
a series of concentric circles drawn about the point charge.
If a given circle does not intersect the wedge,
then the circumference of the circle of radius $R$ will be $2\pi R$.
Once the circle intersects the tip of the wedge (or the apex of the
cone, if we think in 3D), the circumference of the circle will be {\it
smaller}
than $2\pi R$, and the density of field lines radiating outward will be
slightly higher.  Thus the electrostatic energy of the charge is lowered
by moving in farther away from the tip of the cone.  The reverse applies
to a positive disclination, which allows the field to less dense, so that
the electrostatic charge is attracted to the defect.  This effect will 
not affect the vortex-vortex (or charge-charge) interaction if the
particles are much closer to each other than they are to the disclination,
($\vert \Xvec^{(i)} - \Xvec^{(j)}\vert \ll\vert\Xvec^{(i)}\vert,
\,\,\,\, \vert\Xvec^{(j)}\vert $

The vortex-vortex  interactions are
altered by the presence of the defect, introducing terms where the
topological ``charge'' is coupled to two such excitations.  However,
we do not find that three body interactions are introduced among
the vortices themselves.

\subsection{Dislocations:}
\label{subsec:disl}

A similar calculation can be performed for a dislocation 
of Burger's vector $\bvec$ which we choose to be located at the 
origin and corresponding to an additional row of junctions along the negative
x-axis, $\bvec =b\,\, \evec_2$ as in fig.(\ref{defects}.a).
In this case it is easiest to start with the solution for an infinite
half-plane, setting $p=2$ in eq(\ref{cmap}).  
We then deform the half plane into an
semi-infinite slot of width $b$ extending along the positive x-axis
using the
general Schwarz transformation.\cite{panofsky}  
The resulting exact mapping is not invertible in a closed 
form; we approximate it 
with:
\be
z\to z+{b\over 2\pi} \ln{z}
\ee
where we take $|\arg{z}| < \pi$.
This result is
asymptotically correct for $\left\vert{\bvec}\right\vert \ll 
\left\vert{\rvec}\right\vert$, but
lacks the sharp corners of the exact transformation. 
In this limit we find
\bea
E_{\rm disl}&\approx&{1\over 4\pi}
 \sum_{i,j} \rho(\Xvec^{(i)})\,\,\rho(\Xvec^{(j)})\,\,
\ln\left[\left(X_1^{(i)}-X_1^{(j)} -{b\over 2\pi}
(\gamma^{(i)}-\gamma^{(j)})\right)^2
+\left( X_2^{(i)}-X_2^{(j)} + {b \over 2\pi}
\ln{X^{(i)}\over X^{(j)}} \right)\right] 
\nonumber \\
& &
- \sum_i \rho(\Xvec^{(i)})^2\,\,{X_1  b \over 2\pi r^2}
\eea
to lowest order in $\bvec$.  In this expression $\Xvec= X_1\,\,\evec_1 +
X_2\,\, \evec_2$, and the
dislocation is at the origin.  The interaction energy
between a charge and a dislocation at the the point $\xvec$ is in general
\be
E_{\rm disc-vort}\approx {(\Xvec-\xvec) \times \bvec(\xvec) \cdot \evec_3\over 
2 \pi |\Xvec-\xvec|^2}
\ee
to lowest order in $b$.
We see that vortices are attracted or repelled from a dislocation as with a
disclination, but the effect is weaker and depends on the orientation of
the dislocation as well as its distance.
The effect is smallest
when $\bvec$ is perpendicular to the field lines.  

A metric with such a dislocation 
is said to possess {\it torsion}.  Such metrics appear in
general relativity with spinning masses.\cite{K2}

\section{Motion of vortices in a curved space}
\label{sec:motion}

We now examine the actual motion of a single vortex in the above two cases.
Our problem simplifies considerably because we do not have to worry about
vortex-vortex interactions. In order to proceed we need to  specify the metric.  For the
disclination, we choose to work in polar coordinates.  The metric can be
written
\be
g_{ij}=\left(\begin{array}{c  c} 
1 & 0 \\ 0 & \quad (1- {\alpha \over 2\pi})^2r^2 \end{array}\right)\quad
\ee
For $\alpha=0$ (or $p=1$)
 we recover the flat space metric.  Note that the circumference of
a circle of radius $R$ in this space is $(2\pi-\alpha)R$.
 Our Lagrangian in polar coordinates ($r$, $\gamma$) is then  
\bea
L&= &\half \,\, m_{\rm eff} \,\,
g_{ij}\,\,\dot X_i \,\,  \dot X_j
-{\alpha \rho^2\over 4\pi^2}   \,\,\ln{X}\nonumber \\
&=& {m_{\rm eff}\over 2} \left(\dot r^2 + 
(1- {\alpha \over 2\pi})^2 r^2 \dot\gamma^2\right) -
{\rho\over 2\pi} (p-1) \ln{r}
\eea

Solutions of this system can be reduced to quadrature:
\be
(\gamma-\gamma_0) = \int_{r_0}^r {dr'\over \sqrt{c_1 -{c_0^2 p^2 \over r^2}
+ \ln{r}}}
\ee
where $c_1$ and $c_2$ are constants defined by the initial energy and
angular momentum.
In fig.(\ref{disctraj}) we show some sample trajectories for vortices
propagating ballistically in a junction array with  a disclination.
Note that for positive disclinations we see a repulsion, and for negative ones
we see an attraction.  Also shown are trajectories for fictitious ``neutral" 
vortices that propagate along straightlines.  These trajectories show the
effect of the vortex-disclination potential.

For the dislocation with Burger's vector $\bvec = b\,\, \evec_1$, located at
the origin, we choose to work in Cartesian coordinates.
We approximate the metric to lowest order in $b$:
\be
g_{ij}=\left(\begin{array}{c  c} 
1   & - {\dst b y\over\dst 2 \pi r^2}    \\ 
-{\dst b\, y\over\dst 2\pi r^2}& 
\quad 1+{\dst 2 b x\over\dst 2\pi  r^2} \quad
  \end{array}\right)
\ee
In polar coordinates the Lagrangian for the vortex is
\be
L= \half m_{\rm eff} \left[ \dot r^2+r^2\dot\gamma^2 +
2b\dot\gamma\left(\dot r \sin{\gamma} r \dot\gamma\cos{\gamma}\right)
\right] - {b V_0 \over 2\pi} \cos{\gamma}
\ee
For $b=1$ we recover the flat space metric.  
In fig.(\ref{disltraj}) we show some sample trajectories for vortices
propagating ballistically in a junction array with a dislocation. The
trajectory depends upon the relative orientation of the dislocation and the 
vortex.
  Also shown are trajectories for ``neutral" 
vortices that propagate along straightlines.  These trajectories show the
effect of the vortex-dislocation potential.

\section{Conclusion}
\label{sec:conclusion}

In this paper we have shown that the geometry or connectivity of a
lattice can have a strong affect on the dynamics of vortices within
the lattice.  These effects can be seen in the ballistic propagation of
vortices.  Such experiments have already been performed on regular 
lattices.  We propose that similar experiments could be done
with lattices possessing topological defects.

In addition, defects might also change the thermodynamic properties of
the lattice.  The presence of a postitive disclination  acts as an
attractive potential for {\em both} positive and negative vortices,
and it alters the interaction between vortices whose separation is 
comparable to their distance to the defect.  These effects may both alter
the Kosterlitz-Thouless transition within the lattice.  In 
addition, it will alter the interactions in the quantum Hall analogs
proposed in such arrays.\cite{stern,qh}

Since the charge and vortex degrees of freedom are
approximately dual,\cite{fazio,fisher}
we expect similar results for charge excitations within the lattice.
In the limit that we neglect the self-capacitance and the inertial terms in
$K_i$ the two exactly dual.
The interaction potentials for charge excitations will be 
identical to those calculated above, since they are both based on finding
the inverse of the LB operator.   Thus we expect that charge excitations on
the lattice will also interact with distortions of the lattice.  However,
the Coulomb interaction is not strictly two dimensional in real lattices.
We expect therefore that interactions with lattice defects will effectively
be screened out at distance that depends upon the ratio of the nearest
neighbor capacitances to the next-nearest neighbor capacitances.

The motion of charges on twisted lattices also has potential
application in tight-binding models of lattices with dislocations,
or of carbon nanotubes.  In these cases the lattice gives rise
to metrics for the kinetic energy that have non-trivial torsion and
curvature.  These may act as sources of additional scattering or
resistance.

Similar interactions between defects in different ``order parameters'' has
been proposed in other systems.  This idea is at the heart of the
superhexatic state\cite{PRL} in which the defects that mediate the melting
transition interact with superfluid vortex degrees of freedom.  A 
mathematically similar situation arises in the case of tilted 
smectic liquid crystals.\cite{tiltsmec} This system is composed of a 2D
surface of rod-like
molecules that orient themselves at a small
tilt with respect to the surface
normal.   The molecules themselves form a liquid crystal, which has
disclinations as  spatial defects.
The U(1) symmetry associated 
with the orientation of the tilt interacts with the spatial defects in the
smectic.  The interaction energy between the two (sometimes called
``white'' and ``green'' vortices)
is analogous to the vortex-disclination energy we have discussed above.
It is also amusing to note that these Josephson junction
 lattices also provide analogs of how
charges interact in a 2D gravitational system.  

It is quite common to model continuous differential equations by putting
them on a mesh. 
It is usually assumed that the nature of a mesh has no effect on the 
dynamics of the system, if the mesh is sufficiently fine.  We have shown
that topological properties of the mesh can introduce effects in the
dynamics of the system being modelled. Care must be taken when using such
non-integrable metrics, ({\em i.e.} those with torsion or curvature).

In the world of superconducting lattices, distance is
measured in junctions.  This allows the experimenter to create 
lattices where the excitations move in a curved space, producing
several novel effects.

\section*{ACKNOWLEDGMENTS}
This research was supported by Grant No.\ DMR-9502555 from the National
Science Foundation.  Much of this problem was motivated by discussions with
A.~Stern.
We would also like to thank  K.~Burke, S.~M.~Girvin, H.~T.~C.~Stoof and
M.~Wallin 
for stimulating and helpful discussions.

\begin{figure}
\caption{(a)  A square lattice with a dislocation.  It can be viewed as
         a missing horizontal row of islands along the positive x
direction, or an inserted row of island along the negative x-axis.
         (b)  A square lattice with a negative disclination. It is
         formed by removing a $\rm 90^\circ$ wedge from the original
         square array, and then distorting the lattice so that the
         open edges meet. Although the system has a center of 
         three-fold symmetry, note that all sites still have four
         neighbors. To excitations within the lattice it 
         will appear to be a curved space.  
         A positive disclination (not shown)
         is formed by the insertion of a $\rm 90^\circ$ wedge
         of material into the lattice.  It is also possible to 
         form disclinations in triagular lattices by removing or
         inserting a $\rm 60^\circ$ wedge of material.
         \label{defects}}
\end{figure}

\begin{figure}
\caption{A  schematic representation of a junction lattice
where each island is treated as a point. With
each lattice point we
associate two vectors pointing to the nearest neighbors to the left and
above the island.  The vectors $\uvec_{(1)}$ and $\uvec_{(2)}$ may vary as we
move through the lattice.
 \label{lattice}}
\end{figure}

\begin{figure}
\caption{To excitations moving within the a junction array, a lattice  
with a disclination is equivalent to the surface of a cone.
If we slice the cone from the tip out to infinity,
we can flatten it to the plane.  Here we
sketch the field lines of a point charge $P$
in such a case, where the wedge of removed
material forming the disclination
has an opening angle $\alpha$, and is centered at the origin, $O$
(the tip of the cone).
Derivatives of the field are continuous across the cut
so by symmetry,  the field will approach it tangentially.  Because
the field lines are ``excluded'' from the wedge
(the shaded area), the field
lines are forced closer together, increasing the energy.  This
results in a repulsive interaction between the point charge and the
defect.
         \label{wedge}}
\end{figure}
\begin{figure}
\caption{
Sample trajectories for a vortex moving in a disclinated lattice, as in
fig.(1.b).   The disclination is at the origin and 
 has a ``charge'' or missing angle of 
$\pi/4$.  Show here are trajectories for vortices with an initial
velocity $v_0$ and an impact parameter of $r_0$.  The dashed
line is a trajectory for a
 ``neutral'' vortices showing just the effect of geometry. 
The other lines are trajectories of
vortices with interaction energies $\Gamma= E_J/m_{\rm eff} v_0^2$ 
of  1/4,  1/2, and  1, from top to bottom.  The vortices start at the
position $(r_0, -r_0)$.
   The distances are measured in $r_0$.
\label{disctraj}}
\end{figure}
\begin{figure}
\caption{
Sample trajectories for a vortex moving in a lattice with a dislocation,
 as in fig.(1.b).  The dislocation is a missing row of junctions
along the positive x-axis, and is centered at the origin.
Show here are trajectories for vortices with
an initial
velocity $v_0$ and an impact parameter of $r_0$.  Plotted are trajectores
for
``neutral'' vortices (dashed line), and
vortices with interaction energies $\Gamma= E_J/m_{\rm eff} v_0^2$
of 0.40 (middle line), and 1.0 (outer line).   
The distances are measured in $r_0$, and
$b/r_0=0.10$. 
\label{disltraj}}
\end{figure}

\newpage

\typeout{Starting figures...}
\centerline{ {\epsfysize=1.50truein \epsfbox{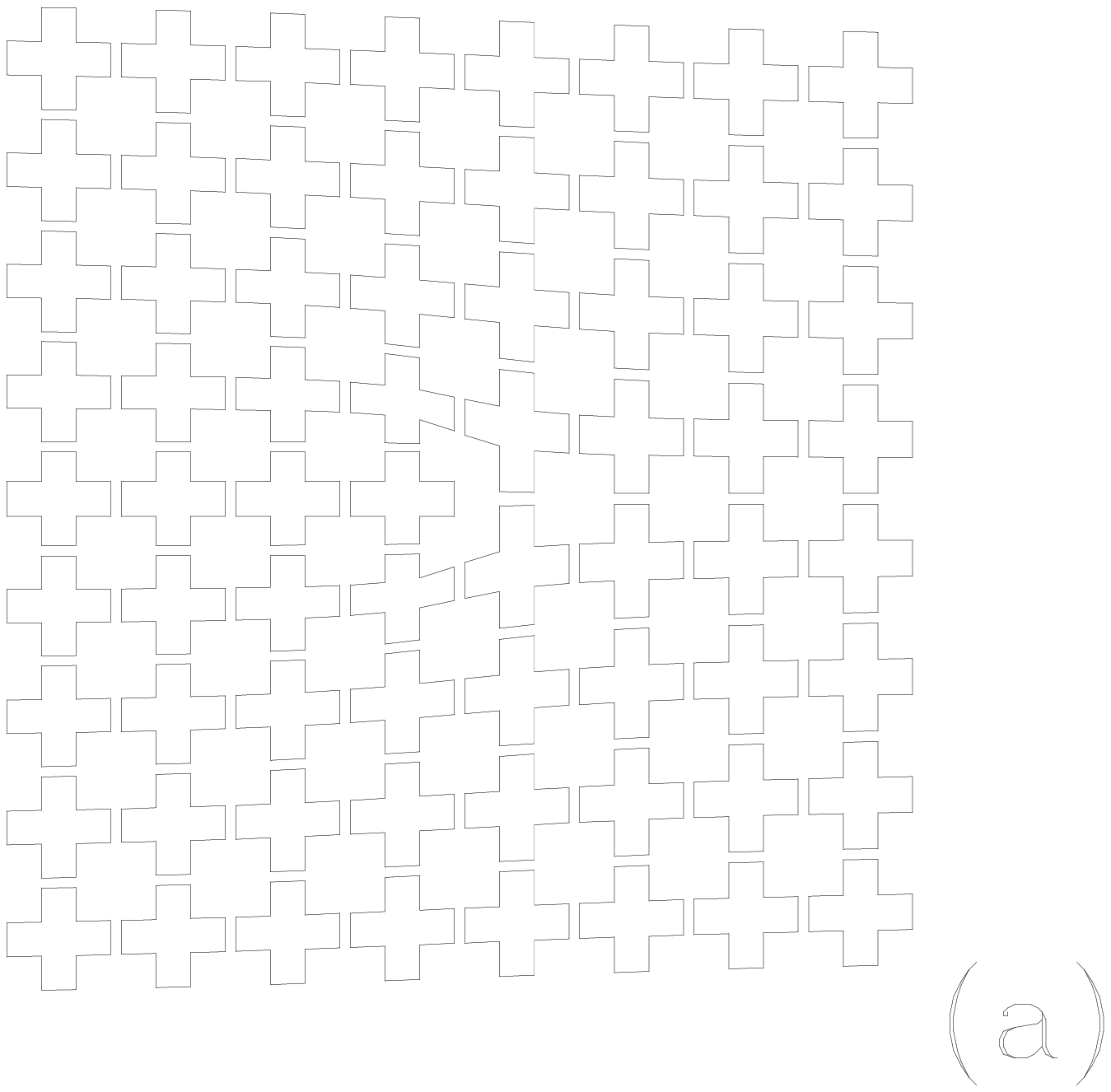}}}
\vfill
\centerline{\Large \bf Figure 1a}
\newpage
\centerline{ {\epsfysize=1.00truein \epsfbox{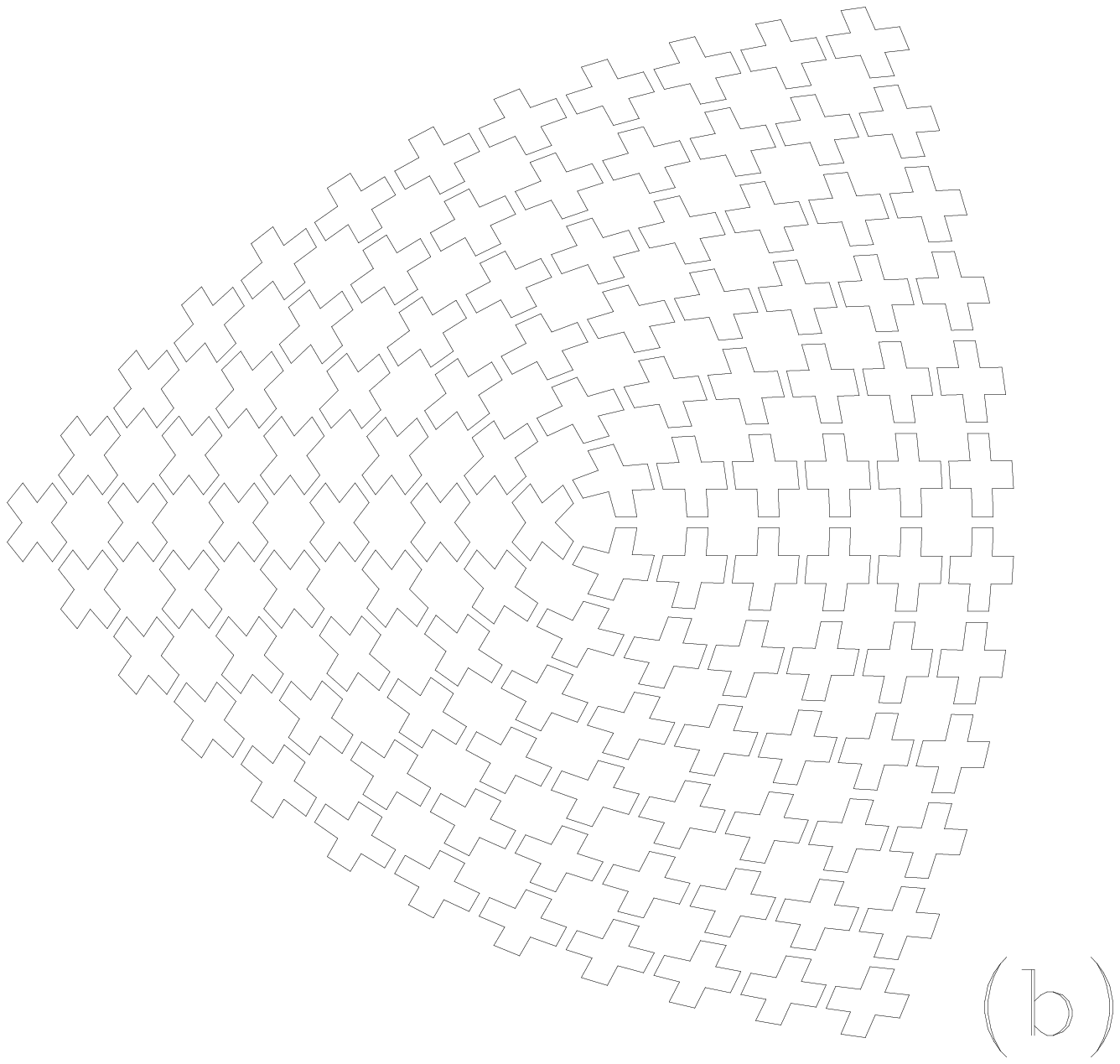}}}
\vfill
\centerline{\Large \bf Figure 1b}
\newpage
\centerline{ {\epsfysize=4.00truein \epsfbox{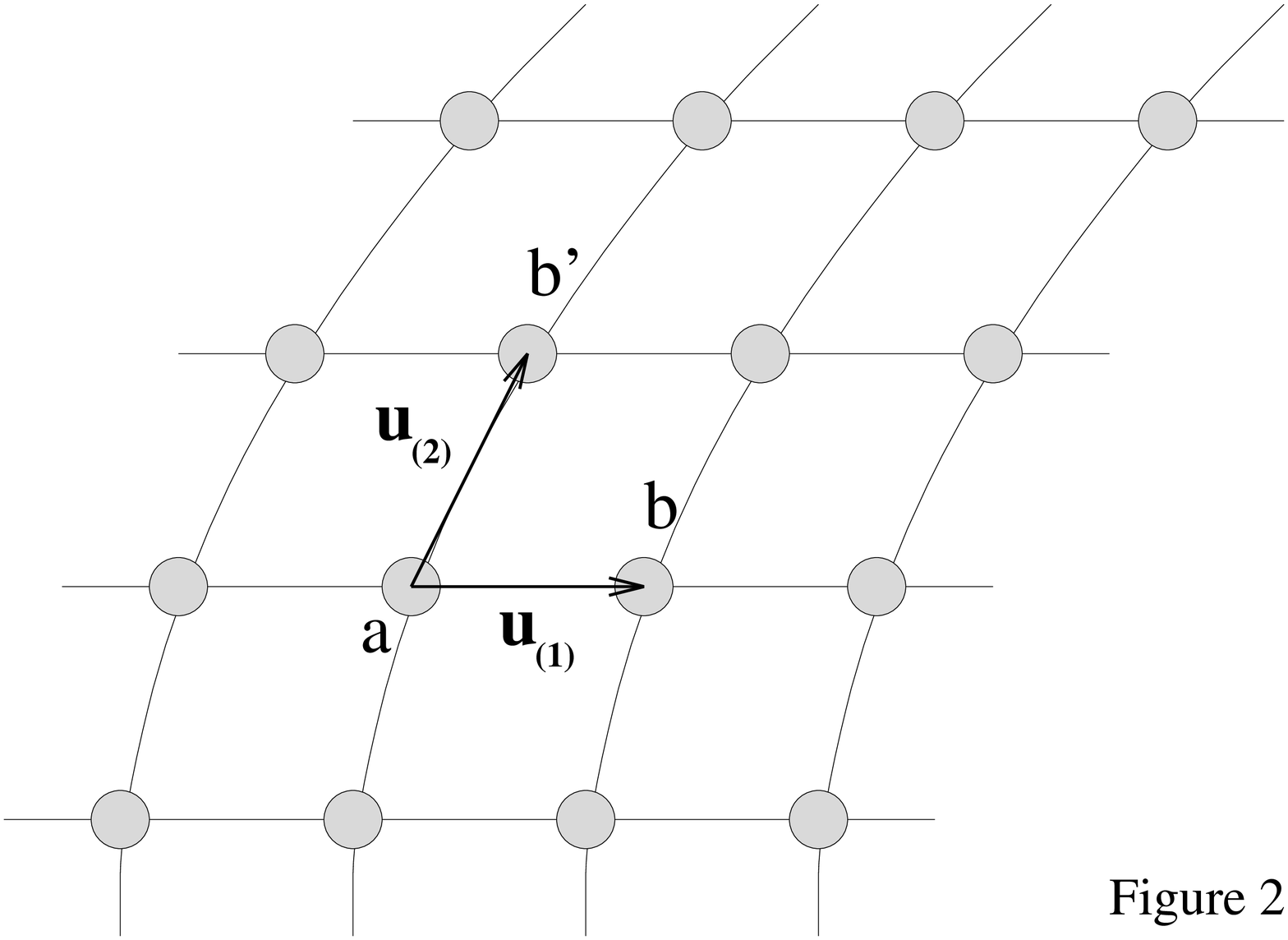}}}
\vfill
\centerline{ {\epsfysize=4.00truein \epsfbox{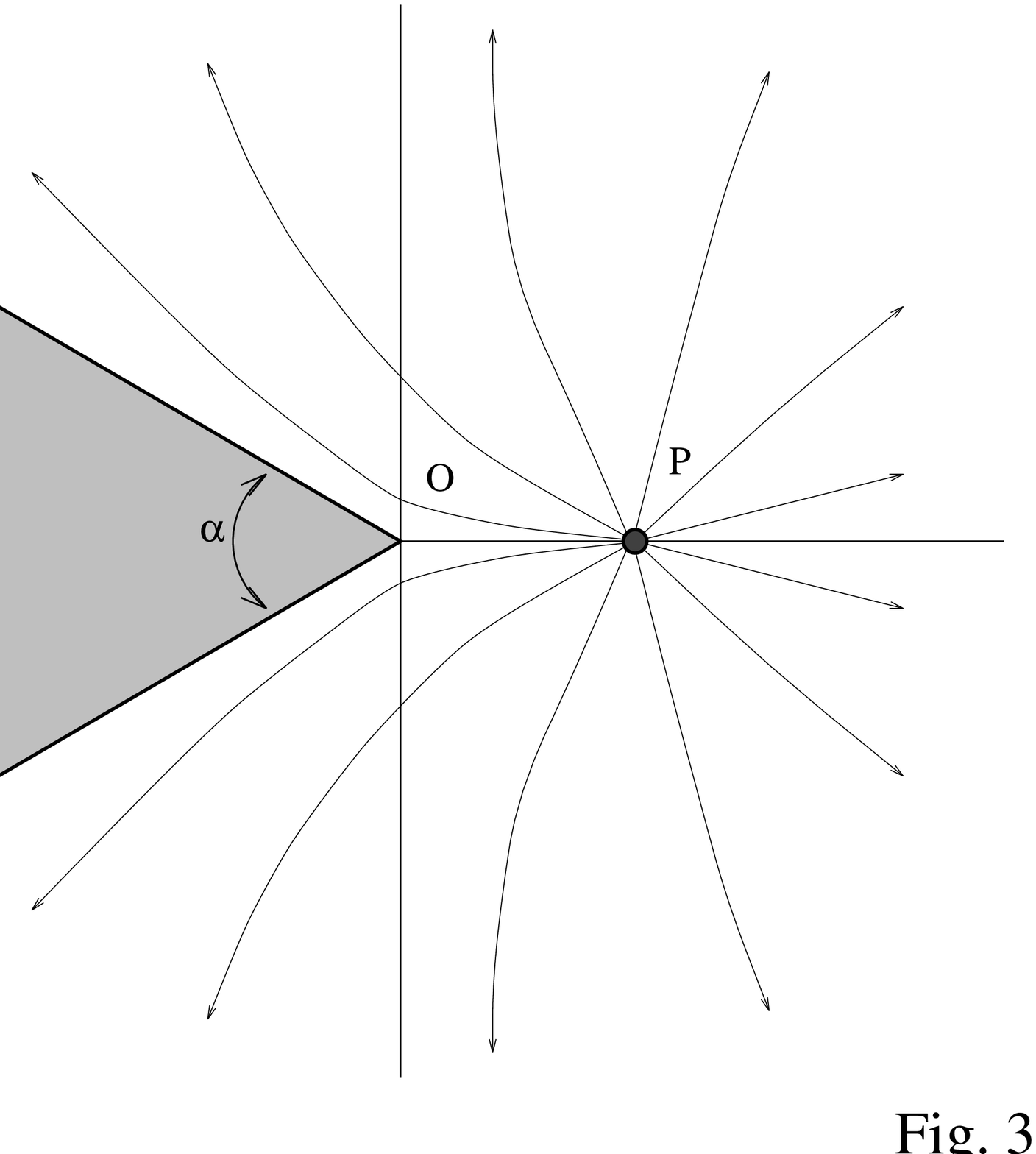}}}
\newpage
\centerline{ {\epsfysize=8.00truein \epsfbox{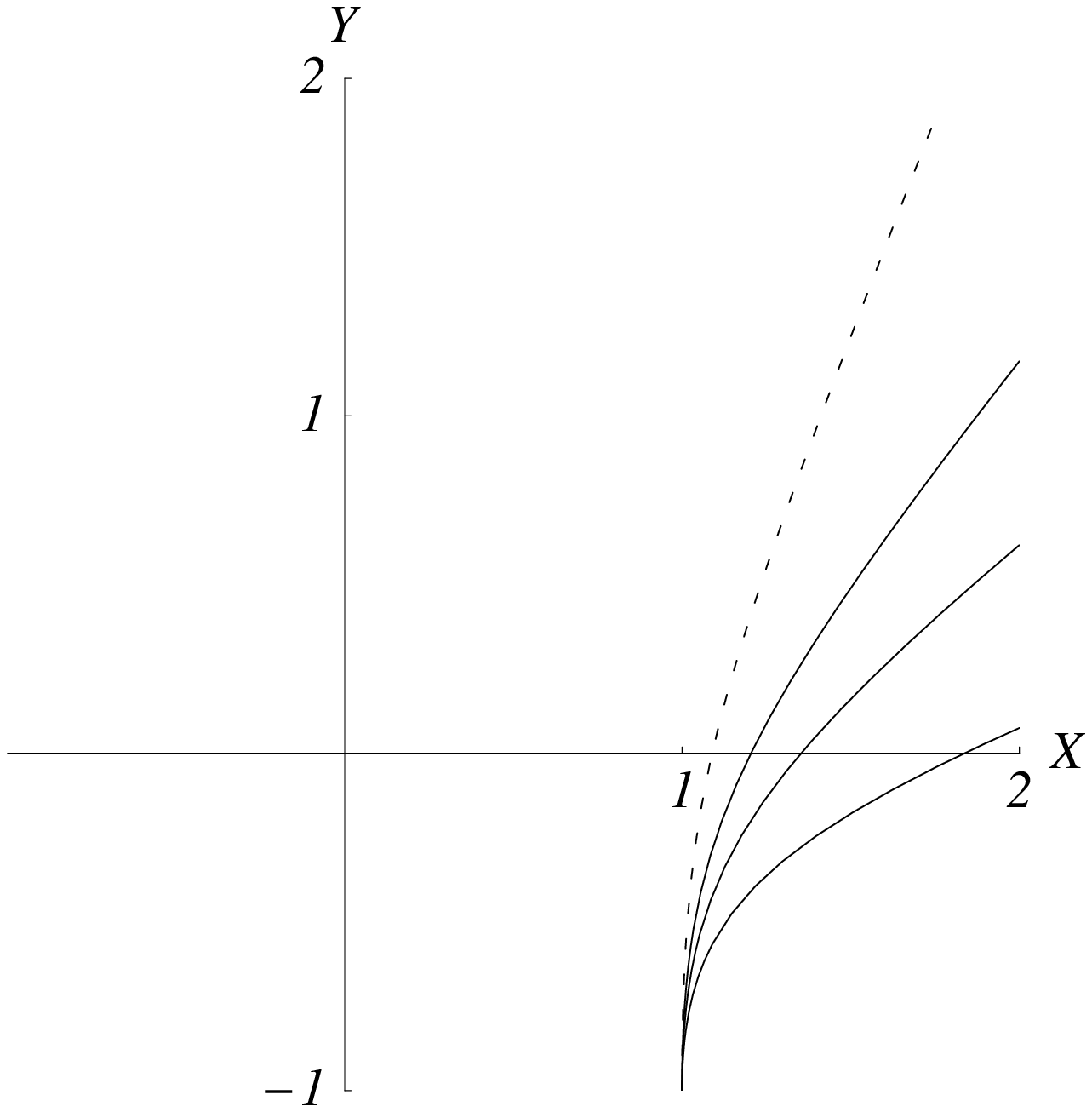}}}
\vfill
\centerline{\Large \bf Figure 4}
\newpage
\centerline{ {\epsfysize=8.00truein \epsfbox{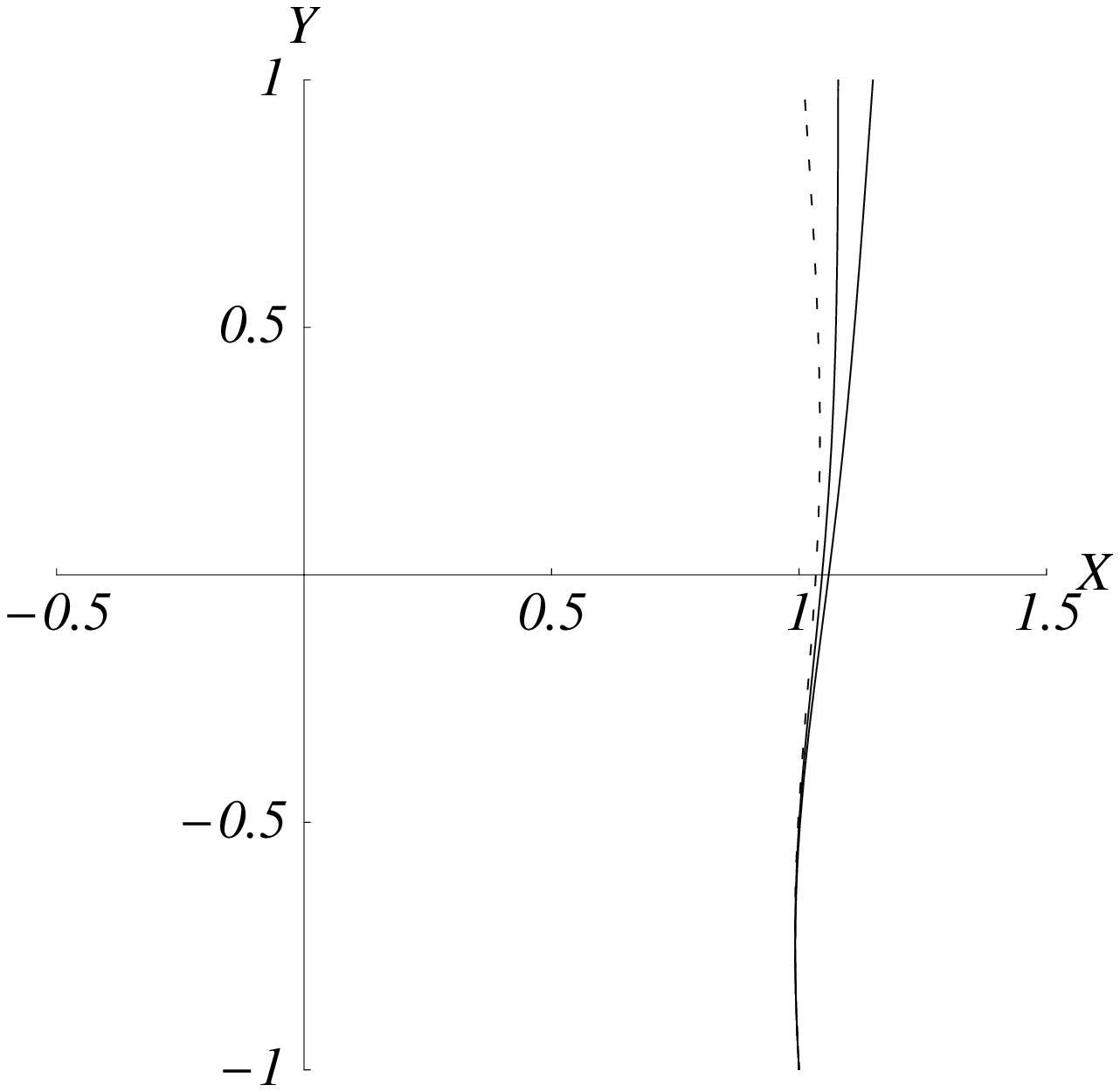}}}
\vfill
\centerline{\Large \bf Figure 5}
\newpage

\end{document}